\pdfoutput=1
\documentclass[twocolumn,showpacs,preprintnumbers,amsmath,amssymb,aps]{revtex4} %si revtex4
\usepackage{graphicx}
\usepackage{color}
\usepackage[all]{xy}
\def\setR{\mathbb{R}}

\newcommand{\norm}[1]{\parallel\!#1\!\parallel} 
\newcommand{\ket}[1]{\mid\nolinebreak\!\!#1\,\rangle} 
\newcommand{\bra}[1]{\langle\,#1\nolinebreak\!\!\mid}

\newcommand{\dron}[3]{\frac{\partial^{#1} {#2}}{\partial{#3}^{#1}}}

\newcommand{\sss}[1]{\scriptscriptstyle #1}

\begin{document}
%\preprint{APS/123-QED}

\title{Two-point function for the Maxwell field in flat Robertson-Walker spacetimes}

%\begin{center}
\author{E. Huguet$^1$}
\author{J. Renaud$^2$}
\affiliation{$1$ - Universit\'e Paris Diderot-Paris 7, APC-Astroparticule et Cosmologie (UMR-CNRS 7164), 
Batiment Condorcet, 10 rue Alice Domon et L\'eonie Duquet, F-75205 Paris Cedex 13, France.  \\
$2$ - Universit\'e Paris-Est, APC-Astroparticule et Cosmologie (UMR-CNRS 7164), 
Batiment Condorcet, 10 rue Alice Domon et L\'eonie Duquet, F-75205 Paris Cedex 13, France.
} 
\email{huguet@apc.univ-paris7.fr, jacques.renaud@univ-mlv.fr}

\date{\today}% It is always \today, today,

%\pacs{}% PACS, the Physics and Astronomy
%$\square$ changed to $\square$

\begin{abstract}
We obtain an explicit two-point function for the Maxwell field in flat Roberson-Walker spaces, thanks to a new gauge condition which 
takes the scale factor into account and assume a simple form. The two-point function is found to have the short distance Hadamard behavior.  
\end{abstract}

\maketitle

%\section{Introduction}\label{Sec-Introduction}
\emph{-Introduction.} Despite numerous works on quantum field in curved spacetimes
\cite{ParkerToms},\cite{BirrelDavies} and the importance of  the flat
Robertson-Walker (RW) spacetimes in current cosmology, it seems that, until
recently, the search for a two-point function for the electromagnetic field in such spacetimes has
been under-looked. In this letter we make a proposal to fill this gap.
Precisely, we derive a two-point function (whose explicit expression is given by Eq.(\ref{EQ-2ptRW})) for the electromagnetic field in flat RW 
spacetimes. To this end we use the Gupta-Bleuler  (GB)
quantization procedure and explain why it applies in this context. The quantization is performed in a covariant gauge
(whose explicit expression is given by Eq.(\ref{EQ-WGaugeRW})) which reduces
to the Lorenz gauge condition in a Minkowski space. 

Two-point functions are of central importance in quantum field theory. In curved spacetimes, explicit expressions are known in a
number of cases. For maximally symmetric spaces, general expressions have been given for both the scalar and the vector field \cite{AllenJacobson}. 
The propagator of the graviton is subject of continuous works especially in de Sitter and Anti-de Sitter spaces 
(see for instance \cite{MoraTsamisWoodard}-\cite{Faizal} 
for recent works). Nevertheless, an explicit expression for the two-point function of the (quantum) Maxwell field (the ``photon's propagator'') in 
RW spaces is seemingly missing.
A recent proposal in conformally flat spacetimes has been made \cite{Faci} in which the electromagnetic field 
appears as a part of a six-dimensional SO$_0(2,4)$-invariant field in a conformally invariant gauge. 
A quantization using the Dirac's procedure for constrained systems has also been recently proposed in flat RW spacetimes \cite{Vollick}. 
But no explicit four dimensional two-point function appears in these two works. In the second one, the choice of the Dirac's method is mainly motivated 
by the alleged inapplicability of the GB condition in cosmological spacetimes. This is related to the  fact that this condition makes explicit use of the 
annihilation operators, while it is well known that there is no preferred vacuum state in a general curved space. This has been 
the starting point of a discussion about the possible contribution of scalar photons to the dark energy \cite{BeltranJimenezMaroto}.
However, following Parker \cite{Parker} the situation of a flat RW space appears as an exception~: 
since the Maxwell equations are conformally invariant, the choice of a preferred vacuum state, the so-called conformal vacuum, 
is possible in such conformally flat spacetimes. We will see that the GB condition makes sense in this context.

The quantization method introduced by Gupta and Bleuler  was designed 
to the electromagnetic field in the Lorenz gauge in the Minkowski space. In this original application, the classical 
Maxwell equations are replaced by others equations  ($\partial^2A_\mu=0$) which provide the modes 
used in this explicit canonical quantization. Unfortunately, 
this process, using the Lorenz gauge in more general spacetimes, yields equations which are often intractable in practice.
This is precisely the situation in flat RW spacetimes. Fortunately, the GB process can be adapted to others gauges
than the Lorenz one. The difficulty can then be circumvented thanks to the conformal
relation between the flat RW space and its global Minkowskian chart. That is, the chart in which the metric is conformal to the Minkowskian 
metric $\eta = \mathrm{diag}(+,-,-,-) $. 
The point is that the conformal map allows us to choose a new gauge condition, in place of the Lorenz gauge on the flat RW manifold, which is conformally
mapped to the Lorenz gauge in the Minkowskian chart. Then, thanks to the conformal invariance of the Maxwell equations the problem 
in the Minkowskian chart is just that of the historical GB method.

%\section{A view of the GB quantization method}\label{Sec-GBReview}
\emph{-A view of the GB quantization scheme.}
In this paragraph we give a concise and practical view of the usual GB quantization method focusing on the electromagnetic field. 
This quantization, historically associated with the Lorenz gauge, is actually more general and   can be,  in particular,  applied in other gauges.
Essentially, the GB method can be viewed as an 
algorithm with the following steps  (commented hereafter):

\noindent1) Define a scalar product on the space of the solutions of the field equations. This product is degenerate if gauge freedom is present.\\
\noindent2) Extend the space of solutions (by considering a new  field equations) in order to eliminate the degeneracy of the scalar product.\\ 
\noindent3)Apply canonical quantization to the field which is solution of the extended equations.\\
\noindent4) Select the subspace of physical states, which correspond to the initial (not extended) field equations using the so-called GB condition.

Note that only the second and the fourth points are specific to the GB scheme. The others belong to the standard canonical 
quantization, which can be presented in a number of equivalent ways.

In the first step, we consider a general bosonic field $A$ on a globally hyperbolic curved spacetime.
The field is assumed to satisfy the Euler-Lagrange equations coming from some quadratic Lagrangian $L$. 
Then, a natural hermitian sesquilinear form on the space of solutions of the Euler-Lagrange equations 
results from the Lagrangian (see for instance Appendix B of \cite{pconf4} for a proof). For two solutions $A$ and $B$ it reads 
\begin{equation}\label{EQ-def_ps}
\langle A, B \rangle=-i\int_\Sigma \sigma_\mu \mathcal{J}^\mu(A,B^*),
\end{equation}
where $\Sigma$ is a Cauchy surface, and $\mathcal{J}^\mu$ is the divergenceless current corresponding to $L$. 
For the free Maxwell field it reads: 
\begin{equation}\label{EQ-def-J}
\mathcal{J}^\mu(A,B) :=  A_\nu  \dron{}{L}{\,\nabla_\mu B_\nu} - \dron{}{L}{\,\nabla_\mu A_\nu} B_\nu.
\end{equation}
Note that   the general expression (\ref{EQ-def_ps}) is for the free scalar field nothing but 
the Klein-Gordon scalar product. In the case of the Maxwell field the gauge invariance makes the above 
scalar product degenerate. This is due to the gauge solutions which are orthogonal to all solutions including themselves. 
With no additional consideration, this property prohibits the canonical quantization. 

The second step is the first part specific to the GB procedure. It amounts to find equations $\mathcal{E}A=0$ which are the 
Euler-Lagrange equations derived from 
a quadratic Lagrangian $L'$ and which satisfy the two following conditions.
First,  the space of solutions of $\mathcal{E}A = 0$ can be equipped with a non degenerate hermitian product.
Second, these equations together with some constraint $\mathcal{C}A=0$ are equivalent to the Maxwell equations $\mathcal{M} A=0$ 
together with a (most frequently covariant) gauge condition $\mathcal{G} A = 0$. 
Thus, the spaces of solutions of $\mathcal{E}A = 0$ and  $\mathcal{M} A=0$ coincide on the subsets defined by their 
respective constraint:
\begin{equation}\label{EQ-Equiv-EQ+Cond}
(\mathcal{E}A=0 \mbox{ and } \mathcal{C}A=0)\Leftrightarrow(\mathcal{M}A=0\mbox{ and } \mathcal{G}A=0).
\end{equation}
In the historical case, for instance, one has $\mathcal{E}A_\mu= \partial^2 A_\mu$, 
${\displaystyle\mathcal{C}A=\mathcal{G}A=\partial\cdot A}$ and $\mathcal{M}A_\mu=\partial^2 A_\mu-\partial_\mu(\partial\cdot A)$.

Then, the third step consists in quantizing the field defined through 
the equation $\mathcal{E} A =0$. This part follows the canonical quantization scheme and is not specific to the 
GB method. In curved spacetimes it thus shares all the well known difficulties of the quantization. 
In practice, it is performed (when possible) through the following steps (see for instance \cite{BirrelDavies}): first find a basis of modes solutions 
of the equation $\mathcal{E}A = 0$ with respect to the (non-degenerate) scalar product, and then determine (choose) the so-called positive frequency modes. 
From them, one obtains the annihilation (and creation) operators, the vacuum, the quantum 
field, the two-point functions \ldots

Finally, the last step of the  GB scheme is to determine the physical states. These are the states which correspond to the classical positive 
frequency solutions satisfying the constraint $\mathcal{C}A=0$. They can be determined thanks to the so-called GB condition. 
It reads in general:
\begin{equation}\label{EQ-anihilateurs_du_champ}
\mathcal{C}\widehat{A}^{(+)}|\Psi_{\sss A}\rangle=0,
\end{equation}
(see for instance Appendix B of \cite{pconf3} for this straightforward 
generalization of the historical condition $\nabla\widehat{A}^{(+)} \ket{\Psi_{\sss A}}=0$).
In this expression the operator $\widehat{A}^{(+)}$ is the annihilator part of the quantum field $\widehat{A}$ and 
$\ket{\Psi_{\sss A}}$ the state corresponding to the classical solution $A$ (that is $\bra{0}\widehat{A}(x)\ket{\Psi_{\sss A}} = A(x)$).  
Finally, the quantum field $\widehat{A}$ fulfills  (\ref{EQ-Equiv-EQ+Cond}) in the mean
on physical states:
 \begin{equation}\label{EQ-MeanOnPhyStates}
 \langle \Psi_{\sss A_1}|\mathcal{M}\widehat{A}|\Psi_{\sss A_2}\rangle = 0~\mbox{ and }~
\langle \Psi_{\sss A_1}|\mathcal{G}\widehat{A}|\Psi_{\sss A_2}\rangle=0,
 \end{equation}
as soon as $\ket{\Psi_{\sss A_1}}$, $\ket{\Psi_{\sss A_2}}$ are physical states.
%$ \mathcal{C}A_i=0$ for $i=1,2$.

It is worth noting that the well known ambiguity in the definition of a vacuum in curved spacetimes, affects the GB 
method both in the quantization of the field and in the determination of the physical states. Nevertheless, we will argue in the 
following that in the particular case of flat RW spacetime, for which the choice of the so-called conformal vacuum is possible, the GB 
procedure is still applicable. 

%\section{Quantization of the electromagnetic field in flat RW spacetimes}\label{Sec-GBFlatSpaces}
%\subsection{\modif{Lesson from Lorenz gauge quantization: the proper use of the Weyl rescaling}}\label{SubSec-UnsucsessLorenz}
\emph{-Lesson from Lorenz gauge quantization}
First, let us attempt to apply the procedure just described to quantize the Maxwell field in the Lorenz gauge $\nabla\cdot A = 0$ in 
a conformally flat RW space.  We first enlarge the space of solutions of the 
Maxwell equations to that of the equations
\begin{equation}\label{EQ-ERW-LGauge}
\square A_\mu + R^\nu_{~\mu} A_\nu  = 0. 
\end{equation}
These equations can be obtained as usual by adding the gauge term 
$\frac{1}{2}\left(\nabla\cdot A\right)^2$ to the Lagrangian of electromagnetism $\frac{1}{4} F^2$, 
where $F$ is the Faraday field strength tensor. In the Lorenz gauge they coincide with the Maxwell equations. 
We then have to determine a basis of modes solution of
(\ref{EQ-ERW-LGauge}). One may attempt to take  advantage of the conformal flatness of the spacetime. 
To this end let us call Minkowskian  coordinates, 
the global coordinates system in which the RW metric element assumes the form 
\begin{equation}\label{EQ-dsRW}
ds^2 = a^2(\tau)(d\tau^2 - d\boldsymbol{x}^2). 
\end{equation}
In that system of coordinates the Eqs. (\ref{EQ-ERW-LGauge}) reads
\begin{equation}\label{EQ-EMink-WGauge}
\partial^2 A_\mu - W_\mu \partial\cdot A + \left(\partial_\mu - W_\mu\right)W\cdot A  = 0, 
\end{equation}
where $W := \mathrm{d}\ln a^2$. Unfortunately,  we didn't succeed in finding a family of modes solutions of the Eqs.  (\ref{EQ-EMink-WGauge}). 
Consequently, we cannot explicitly complete the quantization process, although quantization is still theoretically possible \cite{FinsterStrohmaier}.

Let us remark that the one-form $W$, which appears in Eq. (\ref{EQ-EMink-WGauge}), is generally defined by $W := \mathrm{d}\ln \Omega^2$, 
for a real positive conformal factor $\Omega(x)$. In the system
of coordinates used in (\ref{EQ-dsRW}) one has $\Omega(x) = a(\tau)$ and the only non-vanishing component of 
$W$ is $W_\tau = 2 \mathcal{H}$, two-times the comobile Hubble factor. Throughout
this letter, we nevertheless keep the general notation $W$ since the derivation of all the expressions 
in which $W$ appears do not make use of the exclusive dependence in $\tau$,
they are still valid for $a:=a(x)$.

%\subsection{What lesson can be learned from this failure}
Now, the above unsuccessful attempt leads us to a practicable route. Indeed, a look at Eqs. (\ref{EQ-EMink-WGauge}) makes obvious that
the conformal flatness do not leads to much simplification here. Let us consider more closely the equations (\ref{EQ-ERW-LGauge}) and 
(\ref{EQ-EMink-WGauge}) by themselves.
 The equations  (\ref{EQ-EMink-WGauge}) are simply (\ref{EQ-ERW-LGauge}) written in the 
Minkowskian system of coordinates. Now, it is straightforward to show that the Eqs. (\ref{EQ-EMink-WGauge}) can also be obtained
by adding to the usual Lagrangian of electromagnetism in Minkowski space the gauge fixing term 
$\frac{1}{2} \left(\partial \cdot A + W\cdot A\right)^2$.
In short, to write the Maxwell equations in Lorenz gauge in the RW space is equivalent to write the Minkowskian Maxwell equations in the gauge
%\begin{equation}\label{EQ-WGaugeMink}
$\partial \cdot A + W\cdot A = 0$. 
%\end{equation}

The point is that to write equations over the RW manifold in the global Minkowskian chart in which the Robertson-Walker metric element is (\ref{EQ-dsRW}), 
is equivalent to perform  a conformal transformation (see for instance \cite{BirrelDavies})
between the RW space and a Minkowski space. That is, a  Weyl rescaling  
between the metric manifolds $(\setR^4, g)$ and $(\setR^4, \eta)$, where $g$ and $\eta$ are respectively the RW and the Minkowskian 
metric $\mathrm{diag}(+,-,-,-)$.  
Under such a rescaling, the Maxwell equations are invariant and a form field solution $A$ is mapped to itself since 
its so-called conformal weight is zero. In fact, the rescaling map  induces a transport 
of all mathematical objects (fields, operators,..) between structures defined on the spacetimes. 
In particular, even the equations which are not conformally invariant can be moved between spaces. With this point of view, to quantize the Maxwell equations 
in Lorenz gauge in RW spacetime  is equivalent to quantize the Maxwell equations (since they are conformally invariant)  in the Weyl rescaled 
gauge $\partial \cdot A + W\cdot A = 0$ %(\ref{EQ-WGaugeMink}) 
in the Minkowski space. 

Finally, the lesson from the unsuccessful Lorenz gauge quantization is that if
one wishes to obtain a two-point function, 
one may recognize that the Lorenz gauge condition  in the RW space is not the
best choice. Since the Maxwell equations in Lorenz gauge in Minkowski space are
well known, it is 
far more convenient to start from a gauge condition in RW space which reads as
the Lorenz gauge in the Minkowskian coordinates (or equivalently which 
is conformally mapped to the Lorenz gauge in the Minkowski space). We take this
approach in the sequel.

%\subsection{The quantization in the $W$-gauge}\label{SubSec-LorenzQuantiz}

\emph{-Quantization in the $W$-gauge}
The Lorenz gauge condition in Minkowski space can be conformally lifted to the RW space where it reads  
\begin{equation}\label{EQ-WGaugeRW}
\nabla \cdot A - W\cdot A= 0,
\end{equation}
or, specializing to the case $a=a(\tau)$: %: $\nabla \cdot A = 2 \mathcal{H} A_0$.
\begin{equation*}
\nabla \cdot A = 2 \mathcal{H} A_\tau.
\end{equation*}

As explained in the previous section, we quantize the Maxwell equations on RW spaces in the above $W$-gauge, because in the Minkowskian 
system it reduces to the historical GB quantization in Minkowski space in Lorenz gauge. From the point of view of conformal transformations, 
this amounts to pull back in RW space the whole structure (enlarged space of solutions, basis of modes, two-point function, \emph{etc.}) involved in the
quantization process. Let us consider this construction step by step.

Due to the gauge invariance, the scalar product obtained from the Lagrangian, over the space of solutions of the Maxwell equations is degenerate. 
Following the GB method one first enlarges the space of solutions. This is done by adding to the Lagrangian 
of electromagnetism the gauge term $\frac{1}{2} \left(\nabla \cdot A - W\cdot A\right)^2$. The Euler equations then read
\begin{equation}\label{EQ-ERW-WGauge}
(M A)_\mu  + (\nabla_\mu + W_\mu)(\nabla - W)\cdot A = 0, 
%\square A_\mu - \nabla_\mu \nabla \cdot A + R^\nu_{~\mu} A_\nu  + (\nabla_\mu + W_\mu)(\nabla - W)\cdot A = 0. 
\end{equation}
where we have set $(M A)_\mu := \square A_\mu - \nabla_\mu \nabla \cdot A + R^\nu_{~\mu} A_\nu$. The space of solutions of (\ref{EQ-ERW-WGauge}) is 
endowed with the scalar product (\ref{EQ-def_ps}) derived from the gauge fixed Lagrangian. 
Inspection of (\ref{EQ-ERW-WGauge}) makes obvious that 
the subset of solutions of (\ref{EQ-ERW-WGauge}) defined through the gauge condition (\ref{EQ-WGaugeRW}) are solution of the Maxwell equations. 

The next step in the quantization procedure is to find a basis of modes for the Eqs. (\ref{EQ-ERW-WGauge}). To this end they 
are expressed in the Minkowskian chart, which is equivalent to perform a Weyl rescaling,
and they become
%\begin{equation}\label{EQ-BoxaMink}
$\partial^2 A_\mu = 0$, 
%\end{equation}
as expected from the consideration of the previous section. Then the modes  are, in the Minkowskian coordinates, the familiar exponentials,  they reads 
\begin{equation}\label{EQ-ModesMink}
\Phi^{(\lambda)}_{k,\mu}:= \epsilon^{(\lambda)}_\mu(k) \frac{1}{(2\pi)^3\sqrt{2\omega_{\boldsymbol{k}}}}\exp{\{- i(\omega_{\boldsymbol{k}} \tau - 
\boldsymbol{k} \cdot \boldsymbol{x}})\},
\end{equation}
with $k^0 = \norm{\boldsymbol{k}} =: \omega_{\boldsymbol{k}}$, the forms $\{\epsilon^{(\lambda)}(k)\}$ being the polarization basis. 

It is crucial to remark that, although (\ref{EQ-ERW-WGauge}) is not conformally invariant, it is the conformal lift of the Minkowskian
equation $\partial^2 A_\mu = 0$. Consequently, the above functions are modes of both equations $\partial^2 A_\mu = 0$ and  (\ref{EQ-ERW-WGauge}).
In addition, these modes are of positive frequency with respect
to the time-like Killing vector field of Minkowski space $\partial_\tau$. Since the RW space is conformally flat, $\partial_\tau$ is also a time-like 
conformal Killing vector of the RW spacetime. The modes (\ref{EQ-ModesMink}) are thus positive frequency with respect to the conformal time which means
that the vacuum they define is the so-called conformal vacuum.

It is known that the $\{\Phi^{(\lambda)}_{k,\mu}\}$ form a basis 
of the space of solutions of $\partial^2 A_\mu = 0$ %(\ref{EQ-BoxaMink}) 
endowed with the indefinite scalar product derived through (\ref{EQ-def_ps}) from 
the Minkowskian Lagrangian $L^{\sss M} = \frac{1}{4} F^2 + \frac{1}{2} \left(\partial \cdot A\right)^2$. They also form a basis of the space of solutions 
of Eqs. (\ref{EQ-ERW-WGauge}) endowed with the indefinite scalar product derived from RW Lagrangian 
$L^{\sss RW} = \frac{1}{4} F^2 + \frac{1}{2} \left(\nabla \cdot A - W\cdot A\right)^2$. Indeed, these 
two spaces of 
solutions are identical. This can be seen as follow. First, since the conformal weight of the electromagnetic field is zero, these spaces of solutions, 
contain the same functions. Then, the scalar products defined on them through (\ref{EQ-def_ps}) are equal. In fact, the conformal relation between 
spacetimes implies that $L^{\sss RW} = a^{-4}  L^{\sss M}$. Consequently, using the definition (\ref{EQ-def-J}) of $\mathcal{J}^\mu$  and again the fact that the 
electromagnetic field is of null conformal weight, one obtains $\mathcal{J}_{\sss RW}^\mu(A,B) = a^{-4}  \mathcal{J}^\mu_{\sss M}(A,B)$.  
Since the surface form $\sigma_\mu$ in (\ref{EQ-def_ps}) scales as $\sqrt{g}$, one has $\sigma_\mu^{\sss RW} = a^{4} \sigma_\mu^{\sss M}$. Finally, taking 
into account that 
a Cauchy surface $\Sigma$ defined in the RW spacetime is also a Cauchy surface in the Minkowski chart, one obtains, through the definition (\ref{EQ-def_ps}),
that $\langle A, B \rangle_{\sss RW} = \langle A, B \rangle_{\sss M}$.

Once a basis of modes solutions is known, the Wightman two-point function can be obtained straightforwardly. If one chooses, as usual, a polarization 
basis such that $\eta^{\mu\nu} \epsilon^{(\lambda)}_\mu \epsilon^{(\rho)}_\nu = \eta^{\lambda\rho}$ 
the two-point function takes the familiar Minkowskian form
\begin{equation}\label{EQ-2ptMink}
 D_{\mu\nu}(x,x') = - \eta_{\mu\nu} D^{(s)}_{\sss M}(x,x').
\end{equation}
In this expression, $x$ and $x'$ denotes two points of the RW spacetime whose Minkowskian Cartesian coordinates are $(\tau, x^i)$ and $(\tau', x'^i)$ and
$D^{(s)}_{\sss M}(x,x')$ is the two-point function for the conformal scalar field in Minkowski space. Using the Weyl rescaling, and taking into account that the
conformal weight of the conformal scalar is $-1$, this expression reads
$D_{\mu\nu}(x,x') = - \eta_{\mu\nu} a(x) D^{(s)}(x,x') a(x')$, where $ D^{(s)}(x,x') = a^{-1}(x) D^{(s)}_{\sss M}(x,x') a^{-1}(x')$ is
the two-point function for the conformal scalar field in RW space. Finally, taking into account the manifest symmetry of $D_{\mu\nu}(x,x')$ in (\ref{EQ-2ptMink}),
one obtains a more intrinsic expression for this two-point function in RW spacetime, namely
{\begin{equation}\label{EQ-2ptRW}
%\begin{split}
 D_{\mu\nu}(x,x') = - \frac{1}{2}\left(\frac{g_{\mu\nu}}{a^2} + \frac{g'_{\mu\nu}}{a'^2}\right) a a' D^{(s)}(x,x'),
%\end{split}
\end{equation}
where for brevity we have denoted by a prime the quantities which has to be evaluated at the point $x'$. The above expression is our central result.

{It is worth noting that this two-point function has the Hadamard behavior. Inspection of (\ref{EQ-2ptRW}) shows that 
the short distance behavior is that of $D^{(s)}(x,x')$. Then since $ D^{(s)}(x,x') = a^{-1}(x) D^{(s)}_{\sss M}(x,x') a^{-1}(x')$ the result 
follows from the fact that $\sigma_{\sss RW}(x,x') \simeq a^2(x) \sigma_{\sss M}(x,x')$ for $x$ close to $x'$, the quantities $\sigma_{\sss RW}$,
 and $\sigma_{\sss M}$ being the half of the square of geodesic length between two points $x$ and $x'$ in their respective spacetimes 
(see for instance\cite{SyngeRG}).

Finally, the basis of modes (\ref{EQ-ModesMink})  also allows us to define the quantum field $\widehat{A}$ in the usual way through the expansion over the modes.  
Then the Fock space is built using the standard procedure. The important point here is that the conformal flatness of the RW spacetime allows, thanks to
the existence of a time-like Killing vector field $\partial_\tau$ in Minkowski space, to unambiguously define positive frequency modes. These modes allows 
in turn to define unambiguously annihilation (and creation) operators and consequently a preferred vacuum state: the conformal vacuum. This is a well 
known result of Parker \cite{Parker}, \cite{ParkerToms}. 

As a consequence, the GB condition (\ref{EQ-anihilateurs_du_champ}) which define the subspace of physical states 
is also meaningful. In the $W$-gauge (\ref{EQ-WGaugeRW}) it reads 
\begin{equation*}%\label{EQ-GuptaBleulerRW}
(\nabla - W) \cdot {A}^{(+)} \ket{\Psi_{\sss A}} = 0,
\end{equation*}
where ${A}^{(+)}$ is the annihilator part of the quantum field. This field fulfills the Maxwell equations together with the
$W$-gauge condition, in the mean on physical
states (\ref{EQ-MeanOnPhyStates}). This is not in agreement with the starting point of \cite{Vollick} and \cite{BeltranJimenezMaroto}. However, 
let us emphasize that the GB condition applies in flat RW spacetimes due to the existence of a conformal vacuum. On the contrary this condition can 
loose its meaning in general curved spacetimes where the definition of the vacuum is ambiguous.

\end{document}